\documentstyle[12pt]{article}
\begin{document}

\newcommand{\bs}{\begin{sloppypar}}
\newcommand{\es}{\end{sloppypar}}
\def\theequation{\arabic{section}.\arabic{equation}}



{\large
\begin{center}
{\Large \bf
{ ESTIMATION OF THE MEAN ENERGY OF MUONS  } \\
{ IN MULTILAYER DETECTORS }
}\\
\vskip 1.0cm
{ T.T.Barnaveli, Yu.G.Werbetsky, I.V.Khaldeeva, N.A.Eristavi. } \\
\vskip 0.3cm
{\it
 {Institute of Physics, Georgian Acad. Sci.,}\\
 {Tamarashvili str. 6, Tbilisi - 380077, Republic of Georgia.}\\
 {(E-mail: bart@physics.iberiapac.ge ).}
} \\
\vskip 1.0cm

{\Large \bf Abstract}\\
\vskip 0.5cm


\quotation
{
\normalsize

The technique of muon mean energy determination in multilayer detectors is
developed. The mean energy is measured  by means of average small bursts $m$
i.e.  the number of electrons and positrons generated by muons in the
detecting layers of device via three basic processes --- creation of
$e^+e^-$ pairs, $\delta$-electrons and bremsestrahlung. The accuracy of the
method is considered.

Key words: muon energy, multilayer detectors.

}
\endquotation
\end{center}

\large

\vskip 1.2cm

\section{Introduction.}

{}~~~~~For investigation of the penetrating component of cosmic
radiation by means of multilayer detectors it is essential usually to
estimate the energy of registered events.  This task in cosmic ray physics
 faces the serious technical obstacles, especially for large detectors and
 for muons of very high energies. However many problems in cosmic ray
 physics can be solved if one restricts to the estimation of the mean energy
 of muons in samples of events of different types, e.g. in muon groups of
 some fixed multiplicity $n$. The method of muon mean energy estimation is
 evaluated below.

This method is based on the analysis of a small bursts accompanying the
passage of penetrating particles through the layers of filter and detecting
elements of device. These bursts are generated by muons in the filter layers
of device via three basic processes --- creation of $e^+e^-$ pairs,
$\delta$-electron emission and bremsestrahlung.  The value of mean burst
$m$ turned out to be very effective parameter to evaluate the mean
energy of penetrating particles in multilayer detectors.

The high efficiency of penetrating particle registration and energy
estimation is due to the circumstances quoted below. Despite the fact
that the muon energy losses caused by the pair production process are
close to those caused by the bremsestrahlung, the cross-section of the
first-mentioned process is essentially higher. It increases with the muon
energy value and at $E\geq 10^{12}$eV the probability of $e^+e^-$ pair
creation in the absorber layer of the optimal thickness is close to the
unit. The thickness of the absorber layers is to be chosen with the account
of the requirement of the minimization of particle number fluctuations in
avalanches.  The calculations \cite{GG} show that at the muon energies of
the order of $10^{12} \div 10^{13}$eV the minimal dispersion
is ensured with the thickness of lead absorber of 10 $t$-units, i.e.
$\sim 5cm$.

The approach stated below becomes most effective for the investigation of
high energy muons by means of visual track method of observation. Visual
observation allows to register even the smallest (single particle) bursts,
to distinguish easily the events by their nature and to handle the rare
events of high multiplicity. The influence of transition effects are simple
to be taken into account since it is easy to separate the particles created
in absorber from those created in the lid and the walls of spark chambers.
The application of the multilayer system of spark chambers to the
evaluation of muon energy was proposed earlier in \cite{BBGM}.

Up to day in our experiments (e.g. \cite{BKSE}) we used the
version of this approach described in \cite{BKE}. In the present work some
further development and verification of the method is proposed.

For the analysis of the approach the really working \cite{BKSE} multilayer
spark detector is considered below. The detector consists of 8 layers of
spark chambers separated by a $5 cm$ thick lead absorbers. The area of each
layer of the detector is $4.8 m^2$. The spark chambers are made of glass and
contain neon of high purity at atmospheric pressure. The electrodes are made
of $1mm$ duraluminium. The detector is located under the rock at the depth of
 $190~ hg\cdot cm^{-2}$. This determines the threshold energy of registered
 muons --- about $35 GeV$. The method proposed is in principle applicable for
 the systems containing the registering elements of any type (e.g.
 Heiger-M{\"u}ller counters, neon tubes etc.)

 In chapter 2 the main parameters used are considered. In chapters 3 and 4
 the principle of muon mean energy estimation in multilayer detectors is
 justificated. Chapter 5 is dedicated to the questions of device
 calibration. In chapter 6 actually the process of muon energy estimation
 is described. The dependence $ E(m)$ is given for the detector
 under consideration.


 \section{The main input parameters.}

\setcounter{equation}{0}

 In the real observations of high and superhigh energy cosmic ray muons the
 following parameters are usually available or easily measurable:

\begin{enumerate}
 \item
 {The mean threshold energy $T=T(x)$ of registered muons. This value
 depends on the thickness $x$ of the filter located over detector (rock,
 water etc.) and its absorbing features.}
 \item
 {The exponent $\beta$ of the  spectrum of accompanying bursts
 $Y(m) \propto m^{-(\beta +1)}$.
 According to \cite{G}, in the case of power form of muon energy spectrum
 $P(E) \propto E^{-(\gamma +1)}$ and with the condition of equilibrium of
 accompanying electromagnetic cascade ( i.e. in the vicinity of avalanche
 maximum) the exponents $\beta$ and $\gamma$ approximately coincide.}
 \item
 {The value of mean burst $m$ (i.e. the mean number of $e^+$ and $e^-$
 accompanying the passage of penetrating particle through the multilayer
 system per one layer).}
 \end{enumerate}

 Note that the value of $m$ in the most of cases is  determinable with
 much less error than $T$ and $\beta$.

 If the differential energetic spectrum of penetrating particles
 registered under the filter is defined in normalized power form

 \begin{equation}
 P(E|x) = \frac{\gamma}{T(x)}\cdot \left( 1+ \frac{E}{T(x)} \right) ^{-(
 \gamma +1)} ,
 \label{1.1}
 \end{equation}
 then their mean energy (under the filter)

 \begin{equation}
 <E>~ =~ \int\limits_0^\infty E\cdot P(E)dE
 \label{1.2}
 \end{equation}
 is equal to

 \begin{equation}
 <E>~ =~ \frac{T(x)}{\gamma -1} , \quad \gamma >1 .
 \label{1.3}
 \end{equation}
 Now, if in (\ref{1.3}) one accepts for valuations $\hat \gamma = \hat
 \beta$ (the conditions of such substitution correctness were given
 above in item 2), it becomes possible to determine the mean energy
 valuation $\hat E$ through the energetic threshold $\hat T$ and the
 exponent $\hat \beta$ of the differential spectrum of small bursts.
 However actually the conditions necessary for such application of
 (\ref{1.3}) are not fulfilled. Even for the single muons the spectrum
 differs from (\ref{1.1}); it is significantly flatter at low energies. On
 the other hand, for some samples of events it turns out that $\beta <1$.
 In this case the mean energy $<E>$ formally is not defined at all.
 However $m$ in these samples is steadily recurring in the consecutive
 experiments, so in any case one can consider the value of mean burst $m$
 as the well defined one. This means that the burst spectrum is also not of
 pure power-type. It gets steeper at the big bursts.


 \section{Determination of mean energy of muons.}

\setcounter{equation}{0}

 The parametrization of muon energy $(E_0)$ spectra in atmosphere (both for
 single muons and for those which are a component of the events of any
 other type) will be taken in the normalized form \cite{M}:

 \begin{equation}
 P(E_0) = \frac{\gamma}{T_0}\cdot \frac{ \left( 1+E_0/T_0 \right)^{-\gamma}
 \cdot \left( 1+ E_0/K_0 \right) ^{-1} }{ _2F_1(1, 1; \gamma +1; 1-T_0/K_0)}
 .
 \label{2.1}
 \end{equation}
 Here $_2F_1$ is the Gaussian hypergeometric function, $T_0$ --- the
 characteristic energy describing flattening of the spectrum at low
 energies (of the order of 1GeV); $K_0$ --- the critical energy
 characterizing the competition of two processes --- decay of pions and
 their interaction with air nuclei at the height of effective generation of
 muons \cite{Z}. Generally speaking, the analogous process for $K$-mesons
 is to be taken into account as well. For this reason in (\ref{2.1}) it is
 taken $K_0 \simeq 100 GeV$ --- the effective critical energy approximately
 describing the evolution of the flux of $\pi$- and $K$-meson mixture
 \cite{M}. For a not very large values of filter thickness $x$ it is
 possible not to take into account the large fluctuations of muon energy
 losses and thus the mean energy losses may be taken in the standard form

 \begin{equation}
 -\frac{dE}{dx} = a + bE,
 \label{2.2}
 \end{equation}
 where $a$ and $b$ are the parameters characterizing the material of the
 filter. In this case the form of the spectrum is conserved through the
 evolution of spectrum down the filter depth \cite{B}, while the
 initial energy $E_0$ of muons (the energy at the entry to the filter) is
 the integral of evolution. It is connected with the muon energy $E$ under
 the filter by the relation

 \begin{equation}
 E_0 = (C+E)e^{bx} - C,
 \label{2.3}
 \end{equation}
 where $C = a/b$ is the parameter (energy dimensioned) characterizing the
 material of filter (for the small depths $x$ and small energies $E$ one
 obtains the usually used approximate expression $E_0 \simeq E + ax$ ).

 The relation (\ref{2.3}) describing the full evolution effect of the
 spectrum with respect to the depth $x$ makes it possible to re-write
 the spectrum of muons (\ref{2.1}) under the filter in the form dependent on
 $E$ and analogous to (\ref{2.1}):

 \begin{equation}
 P(E|x) = \frac{\gamma}{T(x)}\cdot \frac{ \left( 1+E/T(x) \right)^{-\gamma}
 \cdot \left( 1+ E/K(x) \right) ^{-1} }{ _2F_1(1 ,1; \gamma +1; 1-T(x)/K(x))
 } .
 \label{2.4}
 \end{equation}
 Here the functions $T(x)$ and $K(x)$ describe the evolution of the
 characteristic energy $T_0$ and critical energy $K_0$ with the depth $x$ of
 the filter:

 \begin{equation}
 \left.
{
 {T \equiv T(x) = C - (C-T_0)e^{-bx} \simeq T_0 + (a-T_0b)x ; }
 \atop
 {K \equiv K(x) = C - (C-K_0)e^{-bx} \simeq K_0 + (a-K_0b)x . }
}
 \right\}
 \label{2.5}
 \end{equation}

 Further we will accept that the energy spectra in the samples of different
 kinds of events (in particular, differing by the number of muons in muon
 groups) may be described by means of functions of (\ref{2.4})-class. Here
 we take the natural assumption that energetic parameters $T$ and $K$
 specified by the process of absorption in filter and by the $\pi$- and
 $K$-decays at high altitudes are the same for all types of samples of
 events. We assume that the difference of spectra of these samples is
 described completely by the difference of values of their exponents
 $\gamma$. Now, once evaluating the value of $\gamma$ by means of any
 method one can estimate the mean energy of muon by the
 distribution (\ref{2.4}):

 \begin{equation}
 <E(x)>~ =~ \frac{T(x)}{\gamma-1}\cdot \frac{_2F_1(2, 1; \gamma +1; Z)
 }{_2F_1(1, 1; \gamma +1; Z) } ,
 \label{2.6}
 \end{equation}
 where

 \begin{equation}
 Z \equiv Z(x) = 1-T(x)/K(x) .
 \label{2.7}
 \end{equation}


 \section{Evaluation of the effective exponent of the spectrum.}

\setcounter{equation}{0}

 To evaluate the effective value of $\gamma$ we use the value of mean burst
 $m$ which is easy to obtain for the given sample of events. On the
 basis of \cite{GG,G62} the mean electromagnetic cascade accompanying the
 passage of muon of fixed energy $E$ through the layer of the absorber of
 the optimal thickness was numerically calculated. The obtained
 $E$-dependence of $\bar m$ can be approximated as

 \begin{equation}
 \bar m(E) = B\cdot \left( E/\epsilon \right) ^\alpha ,
 \label{3.1}
 \end{equation}
 where $\bar m$ is an average burst at fixed energy $E$ of the muon;
 $\epsilon \equiv 1GeV$ --- is the energy scale; the parameters $\alpha, B$
 are determined by the properties and the size of the absorber in the
 detector.  The exponent $\alpha$ one can estimate through the numerical
 calculations with sufficiently high accuracy. The value of the
 normalization parameter $B$ one has to estimate through calibration of the
 device by the well known flux of a single cosmic ray muons (see further).
 Of course, the more accurate results can be achieved by measuring the
 relation like (\ref{3.1}) with a monochromatic beam of muons from
 accelerator at a different energies.  However, only the first way is
 practically available in cosmic ray experiments.

 Averaging the approximate relation (\ref{3.1}) for the $\bar m$ value over
 the spectrum $P(E|x)$ of muon energies (\ref{2.4}) one obtains for the
 twice-averaged burst to be used further again:

 \begin{eqnarray}
  m \equiv <\bar m(E)>~ =~ \int\limits _0^\infty \bar m(E')\cdot P(E'|x) dE'
 =
\nonumber \\
 = B\cdot \left( \frac{T}{\epsilon} \right) ^\alpha \cdot
 \frac{\Gamma (\alpha +1)\Gamma (\gamma -\alpha )}{\Gamma (\gamma )} \cdot
 \frac{ _2F_1(1, \alpha +1; \gamma +1; Z)}{ _2F_1(1, 1; \gamma +1; Z)} ,
 \label{3.2}
 \end{eqnarray}
 or, writing this relation in different way

 \begin{eqnarray}
 \Phi (\gamma ,  m|\alpha ,B,T,K) \equiv
\nonumber \\
\equiv \frac{B}{m}\cdot \left( \frac{T}{\epsilon} \right) ^\alpha \cdot
 \frac{\Gamma (\alpha +1)\Gamma (\gamma -\alpha )}{\Gamma (\gamma )} \cdot
 \frac{ _2F_1(1, \alpha +1; \gamma +1; Z)}{ _2F_1(1, 1; \gamma +1; Z)} -1 =
 0.
 \label{3.3}
 \end{eqnarray}

 Equation (\ref{3.3}) defines the implicit function $\gamma (m|\alpha
 ,B,T,K)$. This equation is to be solved numerically  preliminarily being
 brought to the form convenient for the iteration method:

 \begin{equation}
 \gamma = 1 + G\cdot
 \frac{\Gamma (\gamma - \alpha)}{\Gamma (\gamma -1)}
 \cdot \frac{ _2F_1(1, \alpha +1; \gamma +1; Z)}{ _2F_1(1, 1; \gamma +1; Z)},
 \quad \gamma > 1.
 \label{3.3a}
 \end{equation}
 Here the value

 \begin{equation}
 G \equiv \frac{B}{m} \cdot \left( \frac{T}{\epsilon}\right) ^\alpha
 \cdot \Gamma (\alpha +1)
 \label{3.4}
 \end{equation}
 contains the full experimental information on the sample under
 investigation. The properties of the detector are reflected both in value
 of $G$ and in the parameters $\alpha$ and $Z$ (see (\ref{2.7})). The
 iteration process (\ref{3.3a}) converges rapidly. It is convenient to start
 with the value $\gamma = 2.0$.

 The exponent $\gamma$ evaluated from (\ref{3.3a}) being inserted in
 (\ref{2.6}) defines the searched quantity of the single muon mean energy
 in the sample under investigation.

 So the task is naturally divided into two stages. The first one is
 calibration of the device, i.e. the procedure of evaluation of the
 basic parameters ($\alpha , B,T$ and$K$). The second step is
 application of calibrated device actually to the main task --- estimation
 of separate muon mean energy in the sample of penetrating events.


 \section{Calibration of device.}

\setcounter{equation}{0}

 As one can easily see from above, the goal of our calculations --- a pair
 of quantities $\gamma$ and $E \equiv <E>$ --- is completely fixed by the
 measurable quantity $m \equiv <\bar m>$ and by the set of device
 parameters represented below by the algebraic vector $f \equiv (\alpha
 ,B,T,K)$.  Further the description of the method will be accompanied by
 numerical illustration applied to our installation (see "Introduction").

 Parameter $\alpha$ in (\ref{3.1}) is obtained through calculations and does
 not depend on other parameters. Parameters $T$ and $K$ are obtained from
 (\ref{2.5}).  They do correlate strongly and are defined by properties of
 the filter above the installation and the values of $T_0$ and $K_0$
 quantities for atmospheric muons.

 The parameter of proportionality $B$ is estimated from the precise value of
 measured mean burst $m_1$ in events with single cosmic ray muons.
 The spectrum (\ref{2.1}) of single  muons in atmosphere is well known
 \cite{M} and has the exponent value $\hat\gamma_1 = 2.65 \pm 0.05$. The
 calibration measurements of mean accompanying burst valuation $\hat m_1$
 for the threshold energy $\hat T = 35 \pm 2~GeV$ of registered single
 muons have been carried out (the filter thickness is known with accuracy of
 $\simeq 5 \%$). For the sample of 150 000 single cosmic ray muons the value
 of mean burst for the  considered detector appeared to be $\hat m_1 = 0.173
 \pm 0.002$ (the error is statistical only).

 From (\ref{3.2}) it follows:

 \begin{eqnarray}
 B \equiv
 B(m_1,\gamma_1,\alpha ,T,K) =
\nonumber \\
= m_1\left( \frac{T}{\epsilon} \right) ^{-\alpha} \cdot
 \frac{\Gamma (\gamma_1)}{\Gamma (\alpha +1)\Gamma (\gamma_1 -\alpha )}
 \cdot
 \frac{ _2F_1(1,1, \gamma_1 +1, Z)}{ _2F_1(1,\alpha +1, \gamma_1 +1, Z)} .
 \label{4.1}
 \end{eqnarray}
 Thus the vector of installation parameters $f(\alpha, B, T, K)$ is to be
 calculated through the vector of initial parameters

 \begin{equation}
 p \equiv (b,C,x,T_0,K_0,\alpha ,\gamma_1, m_1),
 \label{4.2}
 \end{equation}
 which were defined above.

 Since all the eight parameters of $p$ vector are to be defined from
 independent experiments it seems to be reasonable to accept the approach to
 these valuations as to normally distributed independent random
 quantities.  For the particular case of our installation we accept:

 \begin{equation}
 \hat p = \left\{ \begin{array}{ccccll}
 p_1: & \hat b   & = & (4.0 \pm 0.4)\cdot 10^{-6} & cm^2\cdot g^{-1} &
 \cite{H}
 \\
 p_2: & \hat C   & = & 460.0 \pm 46.2             & GeV              &
 \cite{H}
 \\
 p_3: & \hat x   & = & 19 000.0 \pm 950.0         & g\cdot cm^{-2}   &
 \cite{BKSE}
 \\
 p_4: & \hat T_0 & = & 1.50 \pm 0.15              & GeV              &
 \cite{M}
 \\
 p_5: & \hat K_0 & = & 110.0 \pm 5.50             & GeV              &
 \cite{M}
 \\
 p_6: & \hat \alpha & = & 0.90\pm 0.05            &                  &
 \\
 p_7: & \hat \gamma_1 & = & 2.65 \pm 0.05         &                  &
 \cite{M}
 \\
 p_8: & \hat m_1 & = & 0.173 \pm 0.002            &                  &
 \end{array} \right.
 \label{4.3}
 \end{equation}

 The transition from initial parameters $p$ to consolidated parameters
 $f(p)$ for the normally distributed quantities $\hat p$ with covariation
 matrix valuation $\hat V$ (in this case diagonal one) is reduced to
 calculation of valuation

 \begin{equation}
 \hat f_\mu = f_\mu(\hat p) + \frac{1}{2} \sum\limits_{i,j = 1}^8
 \hat V_{ij} \cdot \left. \frac{\partial ^2f_\mu}{\partial p_i\partial p_j}
 \right|_{\hat p} , \quad \mu = 1,2,3,4.
 \label{4.4}
 \end{equation}
 Here a correction proportional to the second derivatives (\ref{4.4}) of
 $f(p)$ is small, so all such corrections are further neglected. The
 covariation matrix of the valuations of consolidated parameters

 \begin{equation}
 \hat D_{\mu\nu} = \sum\limits_{i,j=1}^8 \hat V_{ij} \cdot
 \left. \frac{\partial f_\mu}{\partial p_i} \right|_{\hat p} \cdot
 \left. \frac{\partial f_\nu}{\partial p_j} \right|_{\hat p}
 \label{4.5}
 \end{equation}
 naturally is nondiagonal one. Here at least some of the derivatives above
 are to be calculated numerically. The quantities of valuations of
 consolidated parameters and of matrix $D$ are given in Table 1.

Besides, in Fig.1 the dependence $B(\alpha )$ is given for the device under
 consideration. Here the error for $\alpha$ is taken to be zero. The dashed
 lines are the borders of one standard deviation.

According to (\ref{2.6}) one obtains for cosmic ray single muons at the
threshold energy $\hat T=35~GeV$ the value $\hat E_1 = (29.8 \pm 1.9)~GeV$.
Note that valuation of $\hat E$ at $\hat T=35~GeV$  directly from the
experimental data of highly precise work \cite{AKD} leads to the value
$\hat E'_1 = (32.9 \pm 2.0)~GeV$.


 \section{Evaluation of the exponent $\gamma$ and of the mean energy $E$  of
          muons from the sample.}

\setcounter{equation}{0}

 Let us accept that for a sample of penetrating events the experimentally
determined mean burst $\hat m$ is a normally distributed random quantity.
This is approximately valid for sufficiently large samples. In such case
when one obtains the valuations of $\hat \gamma$ and $\hat E$ through
(\ref{3.3}) and (\ref{2.6}). It is possible to consider them as a
substitution of random variables analogous to (\ref{4.4}) and (\ref{4.5}).
Neglecting the corrections to mean values (as we did above) at calculation
of $\hat \gamma$ we restrict ourself to the solution of equation (\ref{3.3})
(the error of calculation must be smaller than statistical one for
$\hat \gamma$ --- see below).  Namely this value of $\hat \gamma$ we will
use to obtain the valuation of $\hat E$ through the relation (\ref{2.6}).
Since the vector of consolidated parameters $\hat f$ and valuation of mean
burst $\hat m$ are statistically mutually independent, it is easy to obtain
the following formulae for the dispersion of $\hat \gamma$ and for
covariation of $\hat \gamma$ with $\hat f$:

\begin{eqnarray}
 \hat \Delta (\gamma ) = \sum\limits_{\mu,\nu =1}^4 \hat D_{\mu\nu} \cdot
 \left. \frac{\partial \gamma}{\partial f_\mu} \right|_{\hat f,\hat m}\cdot
 \left. \frac{\partial \gamma}{\partial f_\nu} \right|_{\hat f,\hat m} +
\hat \sigma^2_m\cdot \left[ \frac{\partial \gamma}{\partial m} \right]
^2_{\hat f,\hat m} ;
\nonumber \\
 \hat \Delta (\gamma ,f_\rho ) = \sum\limits_{\mu,\nu =1}^4 \hat D_{\mu\nu}
 \cdot \left. \frac{\partial \gamma}{\partial f_\mu} \right|_{\hat f,\hat m}
 \cdot
 \left. \frac{\partial f_\rho}{\partial f_\nu} \right|_{\hat f,\hat m} ;
 \nonumber \\
 \quad \rho = 1,2,3,4.
\label{5.1}
\end{eqnarray}
 Here $\hat \sigma^2_m$ is the valuation of dispersion of the mean burst
 $\hat m$ which is got from the experiment. Derivatives of $\gamma (f,n)$
 are calculated from (\ref{3.3}) as a derivatives of implicit function:

 \begin{eqnarray}
 \frac{\partial \gamma}{\partial f_\nu} = -~\frac{\partial \Phi /\partial
 f_\nu}{\partial \Phi /\partial \gamma}, \nonumber \\
 \frac{\partial \gamma}{\partial m} = -~ \frac{\partial \Phi /\partial
 m}{\partial \Phi /\partial \gamma}, \nonumber \\
 \nu = 1,2,3,4.
 \label{5.2}
 \end{eqnarray}
 The dispersion of mean energy valuation for muons of the sample is obtained
 analogously ($E = E(f,\gamma )$, see (\ref{2.6})):

 \begin{eqnarray}
 \hat\sigma^2_E = \sum\limits_{\mu,\nu =1}^4 \hat D_{\mu\nu} \cdot
 \left. \frac{\partial E}{\partial f_\mu} \right|_{\hat f,\hat \gamma}\cdot
 \left. \frac{\partial E}{\partial f_\nu} \right|_{\hat f,\hat \gamma} +
 \nonumber\\
 +
 2\cdot \sum\limits_{\rho =1}^4 \hat \Delta (\gamma ,f_\rho ) \cdot
 \left. \frac{\partial E}{\partial f_\rho} \right|_{\hat f,\hat \gamma}\cdot
 \left. \frac{\partial E}{\partial \gamma} \right|_{\hat f,\hat \gamma} +
\nonumber\\
 +
 \hat \Delta (\gamma ) \left[ \frac{\partial E}{\partial \gamma} \right] ^2.
 \label{5.3}
 \end{eqnarray}
 For example the sample of 4-muon events with the mean burst $\hat
 m_4 = 0.45\pm 0.07$ leads to the valuations of $\gamma_4$ and $E_4$:

 \begin{equation}
 \hat \gamma_4 = 1.65 \pm 0.13 \quad , \quad \hat E_4 = 91 \pm 21 GeV.
 \label{5.4}
 \end{equation}

 In Fig.2 the set of dependencies $E(m)$ is given for the set of
 values $\alpha = 0.80; \; 0.85; \; 0.90; \; 1.00$ ---  this
 parameter may vary most probably in different devices (if one pais no
 attention to the possibility of soil-thickness variation above the
 installation). Here $\sigma_m = 0, \; \sigma_\alpha = 0$ is accepted. It is
 easy to see that in the region of relatively small energies (bursts) the
 result is not very sensitive to the variations of $\alpha$.

 In Fig.3 the dependencies $\gamma (m)$ and $E(m)$ are given
 for the above quoted values of consolidated parameters $f$ and for
 $\sigma_m = 0$, i.e.  the corridor of standard deviations indicated for
 central values of mean bursts is the systematic error of the
 detector.  Separately standing points on Fig.3 are the following:

\begin{enumerate}
 \item
 {--- direct evaluation of exponent $\hat\beta_4$ in sample of 4-muon
 events for the case of exponential cutoff of the spectrum in the area of
 large bursts.}
\item
{--- value $\hat E_4$, calculated according to
 (\ref{2.6}) for the above-mentioned value of $\hat\beta_4$ by
 identification $\hat\gamma_4 = \hat\beta_4$ (see "Introduction"). }
\item
{--- direct evaluation
 of exponent $\hat\beta '_4$ in sample of 4-muon events for purely power
 spectrum of bursts (argument $m$ is calculated from this
 spectrum).}
\item
{--- value $\hat E'_4$ corresponding to the  equality
$\hat\gamma '_4 =\hat\beta '_4$ is calculated according to (\ref{1.3}).}
\end{enumerate}

 Only statistical errors are given for these four points, as well as for the
 central point, got by the regular method proposed.

Note that evaluation of $E$ is rather stable in respect to variation of
spectrum shape at large bursts. In particular if one takes for spectrum in
the sample the power approximation (\ref{1.1}) but estimates $m$
neglecting the very large bursts, the values of $E$ obtained for close
$m$-s differ insignificantly.


\section{Conclusion.}

\setcounter{equation}{0}

The proposed method for muon mean energy estimation with the help of
accompanying burst mean value $m$ of the sample of penetrating events
is based on the solution of implicit equation (\ref{3.3}) for the spectrum
exponent $\gamma$. The shape of this equation certainly depends on the
class of functions chosen to approximate the spectrum of sample under
investigation.

It is obvious that the obtained evaluations of energy depend also on the
precise definition of the class of events appertained to the sample. E.g.
for the sample of penetrating events of fixed number of muons these
evaluations, in general, depend on the size of installation (or its
sensitive area) in which the registration of fixed number of muons is
demanded to attribute the event to the indicated sample.

With account of these restrictions the described method allows the reliable
estimation of muon mean energies by means of accompanying burst mean value
which can be measured in experiment.

Note that for the nontrack devices the concept of the "mean burst" can be
substituted by e.g. "mean number of changed-state cells", or something like
it, according to applied detecting elements. Of course this will require the
corresponding correction but in principle differs by nothing from the above
said.


\vskip 1.5cm
{\Large \bf Acknowledgments. }
\bigskip

Authors express their cincere gratitude to the technical personnel of
laboratory for dedicated labor which in hard circunstances made it possible
to obtain the experimental data for the calibration of the method.


\newpage
{\Large \bf Figure Captions.}

\begin{itemize}
\item[Fig.1.]
{The calibration parameter $B$ dependence on the exponent $\alpha$ value
for the installation considered.}
\item[Fig.2.]
{The average energy $E$ sensibility to the exponent $\alpha$ variations.
Only systematic erreos are shown.}
\item[Fig.3.]
{The exponent $\gamma$ and average muon energy $E$ dependence on the mean
burst $m$ of the sample.}
\end{itemize}

\newpage

\vskip 4.0cm
\large

\begin{table}

\begin{picture}(600,300)
\setlength{\unitlength}{1.0mm}
\thicklines
\put(0,0){\framebox(156,81)[bl]{
\put(0.5,0.5){\framebox(155,80)[bl]{
\put(15.5,0.5){\line(0,1){80}}
\put(0.5,65.5){\line(1,0){155}}
\put(15.5,40.5){\line(1,0){140}}
\thinlines
\put(15.5,50.5){\line(1,0){140}}
\put(15.5,25.5){\line(1,0){140}}
\put(50.5,0.5){\line(0,1){25}}
\put(85.5,0.5){\line(0,1){25}}
\put(120.5,0.5){\line(0,1){25}}
\put(50.5,40.5){\line(0,1){10}}
\put(85.5,40.5){\line(0,1){10}}
\put(120.5,40.5){\line(0,1){10}}
\put(50.5,65.5){\line(0,1){15}}
\put(85.5,65.5){\line(0,1){15}}
\put(120.5,65.5){\line(0,1){15}}
\put(3.5,77.5){\line(1,-1){9}}
\put(3.5,2.5){\shortstack{$[1]$\\$[2]$\\$[3]$\\$[4]$}}
\put(33,22){\makebox(0,0){$ +2.025\cdot 10^{-3} $}}
\put(33,16.5){\makebox(0,0){$ -7.153\cdot 10^{-5} $}}
\put(36.5,10.5){\makebox(0,0){$ 0 $}}
\put(36.5,5){\makebox(0,0){$ 0 $}}
\put(68,22){\makebox(0,0){$ -7.153\cdot 10^{-5} $}}
\put(68,16.5){\makebox(0,0){$ +2.640\cdot 10^{-6} $}}
\put(68,10.5){\makebox(0,0){$ -4.863\cdot 10^{-4} $}}
\put(68,5){\makebox(0,0){$ -3.572\cdot 10^{-4} $}}
\put(106,22){\makebox(0,0){$ 0 $}}
\put(103,16.5){\makebox(0,0){$ -4.863\cdot 10^{-4} $}}
\put(103,10.5){\makebox(0,0){$ +2.759 $}}
\put(103,5){\makebox(0,0){$ +2.227 $}}
\put(141.5,22){\makebox(0,0){$ 0 $}}
\put(138,16.5){\makebox(0,0){$ -3.572\cdot 10^{-4} $}}
\put(138,10.5){\makebox(0,0){$ +2.227 $}}
\put(138,5){\makebox(0,0){$ 28.43 $}}
%
\put(10.5,75.5){\makebox(0,0){$\nu$}}
\put(5.5,70.5){\makebox(0,0){$\mu$}}
\put(20.0,76.5){\makebox(0,0){$[1]$}}
\put(55.0,76.5){\makebox(0,0){$[2]$}}
\put(90.0,76.5){\makebox(0,0){$[3]$}}
\put(125.0,76.5){\makebox(0,0){$[4]$}}
\put(33.0,70.0){\makebox(0,0){$\alpha$}}
\put(68.0,70.0){\makebox(0,0){$\beta$}}
\put(103.0,70.0){\makebox(0,0){$T, ~GeV$}}
\put(138.0,70.0){\makebox(0,0){$K, ~GeV$}}
%
\put(33.0,45.5){\makebox(0,0){$ 0.90 \pm 0.05 $}}
\put(68.0,45.5){\makebox(0,0){$ 0.0088 \pm 0.0016 $}}
\put(103.0,45.5){\makebox(0,0){$ 35.1 \pm 1.7 $}}
\put(138.0,45.5){\makebox(0,0){$ 135.6 \pm 5.3 $}}
\put(78.0,85.0){\makebox(0,0){Consolidated Characteristics of the Device}}
\put(85.0,58.0){\makebox(0,0){Consolidated Parameters $f_\nu$ of the Device}}
\put(85.0,33.0){\makebox(0,0){ Covariation Matrix $D_{\mu\nu}$ }}
}}
}}
\end{picture}
\caption{}
\end{table}

\end{document}